\documentclass[ DIV=calc, BCOR=25pt, paper=a4, fontsize=10pt, twocolumn]{scrartcl}	 

\usepackage[english]{babel} 
\usepackage[journal=jmcmar,usetitle=false]{achemso}
\usepackage{graphicx}
\usepackage[normal, small, labelfont=bf,up,textfont=up]{caption} 
\usepackage{float}

\title{Mass Spectrometry--An Alternative in Growth~Hormone Measurement\footnote{submitted for publication to {\em Bioanalysis;}\hspace{7pt}$^\pi$Physikalisch-Technische~Bundesanstalt~(PTB), D-38116 Braunschweig, Germany; $^\rho$Institute of Laboratory Medicine, Clinical Chemistry and Molecular Diagnostics, University of Leipzig, D-04103 Leipzig, Germany; \hspace{5pt}$^\sigma$E-mail: andre.henrion@ptb.de} } 

\author{Cristian G. Arsene,$^\pi$  J\"urgen Kratzsch\hspace{1pt}$^\rho$ and Andr\'e Henrion\hspace{1pt}$^{\pi,\sigma}$\hspace{3pt}}

\begin{document}
\maketitle
\thispagestyle{empty}
\abstract{\noindent Growth hormone (GH) constitutes a set of closely related protein isoforms. In clinical practice, the disagreement of test results between commercially available ligand-binding assays is still an ongoing issue, and incomplete knowledge about the particular function of the different forms leaves an uncertainty of what should be the appropriate measurand. Mass spectrometry is promising to be a way forward. Not only is it capable of providing SI-traceable reference values for the calibration of current GH-tests, but it also offers an independent approach to highly reliable mass-selective quantification of individual GH-isoforms. This capability may add to reliability in doping control too. The article points out why and how.} 

\section*{Measurand}
Growth hormone (GH) is a heterogeneous polypeptide consisting of a number of different isoforms and variants\cite{Baumann2009}. These are closely related to one another as originating from the same gene, i.e., information about GH-amino acid sequence stored on DNA. In reality, however, just parts of this sequence are read when assembling the GH while others are skipped (alternative splicing). Typically, therefore, the different forms share much of the parent sequence, but not all.\par 

\begin{figure*}[]
\centering
\includegraphics[width=0.85\textwidth]{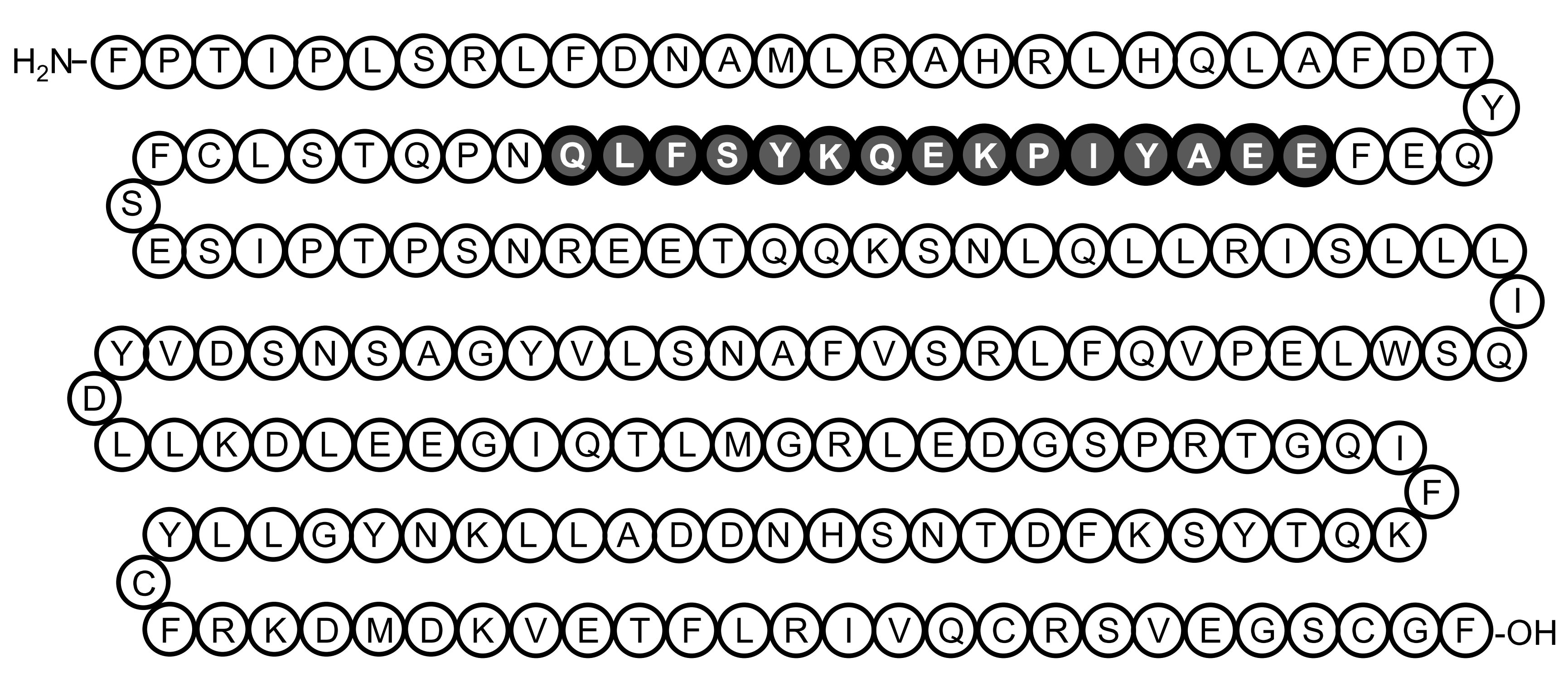}
\caption{Parent-primary structure (amino acid sequence) of GH. With 22~kDa-GH, all 191 amino acids are present, whereas 20~kDa-GH lacks in substretch 32--46 (grey-filled circles).} \label{GHSequence} 
\end{figure*}

The sequence of the main GH-form (22~kDa-GH) is shown in Figure~\ref{GHSequence}. This form accounts for 58--82\% of the circulating GH\cite{Baumann2009}. There is another one, 20~kDa-GH, which is lacking in 15 out of the 191 amino acids of the main form. These are marked by grey-filled circles in the Figure. Next to 20~kDa-GH, which is estimated to make a 3--15\% fraction of GH forms\cite{Baumann2009}, sequence fragments consisting of amino acids 1--43 have been identified and the complementary stretch (44--191) has been detected in blood\cite{Sinha1994}.\par

Further complicating matters, different variants may exist of the same isoform by modification of the side chains, typically resulting from deamidation or glycosylation. Last, but unlikely to complete the list, the hormone has been detected in dimeric or even higher molecular forms\cite{Baumann2009}, e.g. (22kDa)$_2$, (20kDa)$_2$ or (22kDa-20kDa), as protein-protein complexes, besides dimers (22kDa)$_2$  with the monomers linked by disulfide bonds.\par  

Obviously, in this setting, the crucial problem consists in knowing, which are the different functions in regulation, exerted by the different forms, and which of them is (are) correlating best with a particular clinical condition or GH-related disease to be detected or measured. Finding out about this requires the analytical technology to resolve these forms while providing reliable quantitative results for each of them.\par

Mass spectrometry (MS), relying on a fundamentally different principle of analyte-recognition enables evaluation of ligand-binding assays (LBAs) without being subject to the same limitations as these. The analytical selectivity generally inherent to MS can be deployed to discriminate isoform-specific GH-sequence fragments. This, in combination with the option to use isotope-labeled internal standards, makes MS an interesting alternative that can contribute to respond to the challenge stated above.\par
The subsequent discussion is intended to illustrate this fact by the comparison of both kinds of approach, ligand binding vs. mass selective.

\section*{Recognizing by epitope}

\subsection*{Scope and bounds}

For a long time, the ligand-binding assays have been the only way to quantification of GH in biological samples. Apart from this inevitableness, ease of practical implementation and automation, as well as effectiveness in time and costs per analysis are assets of the method principle. Importantly also, sensitivity requirements of clinical tests can be satisfied by use of appropriate detection techniques, as e.g. chemiluminescence. Then, detection limits can be achieved of 0.2 ng/mL, or less\cite{Bidlingmaier2010} as e.g. needed for measurement of GH-concentrations in patients with growth hormone deficiency (GHD).\par

\begin{figure}[H]
\hfill
\includegraphics[width=0.4\textwidth]{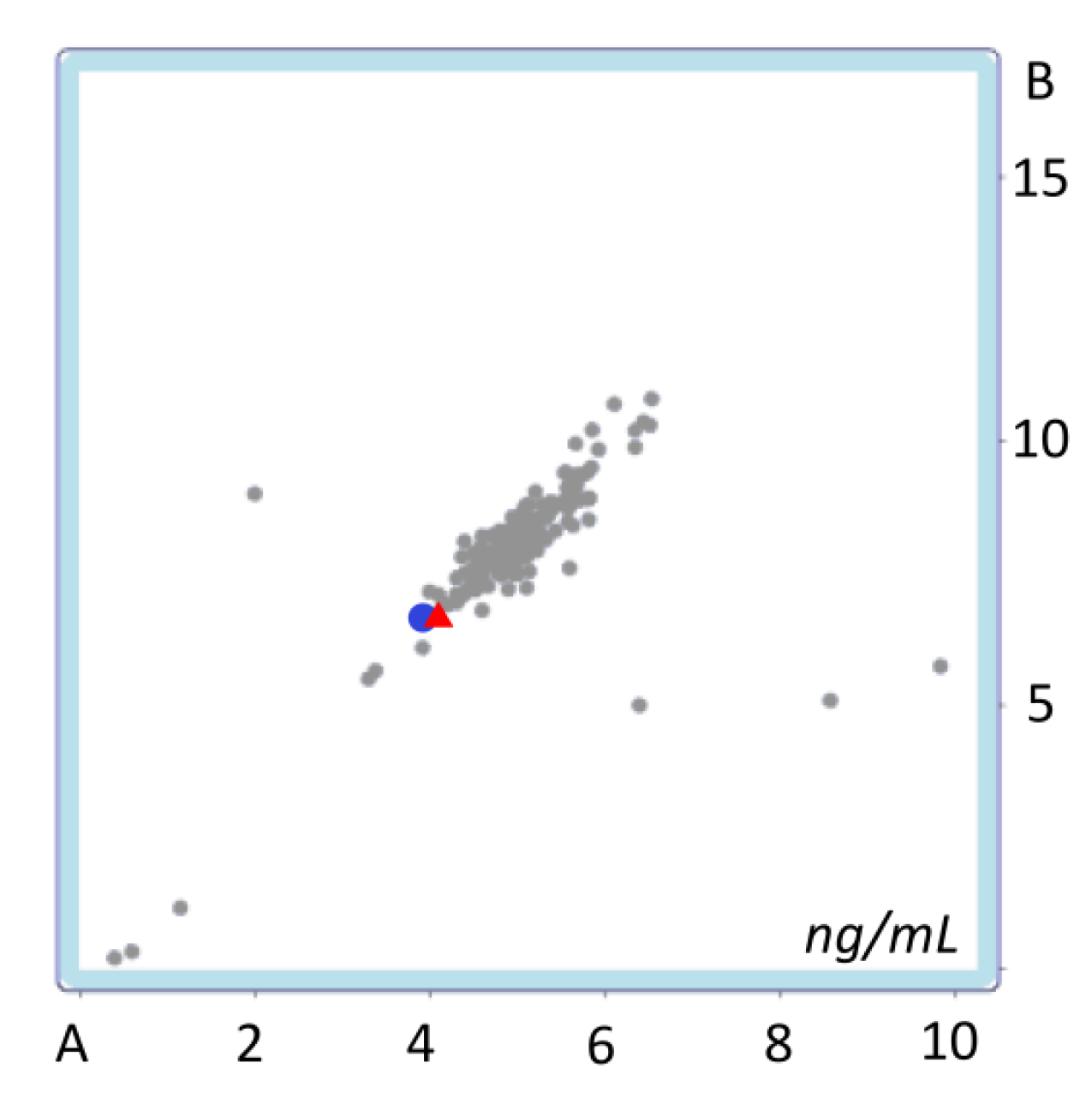}
\caption{Results in an intercomparison of testing laboratories for the measurand 'growth hormone' (HP 4/12) in 2012. A number of 14 different test kits had been employed by 186 participating laboratories. Each laboratory ran two samples (serum pools A and B) under their routine conditions. Pairs of results (one for pool A, one for B) make one dot in the picture. Red triangle and blue circle: MS results (red: T6-based, blue: T12-based). (Reproduced/adapted with permission from https://www.dgkl-rfb.de.  \copyright~2012 DGKL, Referenzinstitut f\"ur Bioanalytik.)} \label{RfB} 
\end{figure}

At the same time, the limited quality of GH-data is still hampering the reliability of conclusions in diagnostic practice. Figure~\ref{RfB} exemplifies this with results reported by testing laboratories in one of the periodic studies organized by RfB DGKL, one of the German providers of External Quality Assessment Services (EQAS). With both of the serum pools, A and B, the spread of data covers a considerable portion of the range of values typically found in the population as a whole, in spite of the fact, that A and B represent serum of just a single subject, each. Decision limits for diagnostic GH-stimulation tests, for illustration, are in the range of about 4--10~ng/mL\cite{Mueller2011}.\par

The situation very much resembles what is observed in other countries too. Some further examples describing the impact of solely the choice of the assay-method (laboratory) employed, on patient classification, are provided by \citeauthor{Wieringa2014}\cite{Wieringa2014} In one of them, a factor of 6, on average, was in-between the results of two different test-kits that had been used with an oral glucose tolerance test\cite{Arafat2008}.

\subsection*{Limiting factors}

A recent review\cite{Wieringa2014} summarizes the most likely causes for intermethod-discrepancies, and why they continue to be unsatisfactory in spite of improvements resulting from standardization efforts during these last ten years.\par

One category of sources is the different affinity of antibodies toward different forms of GH. In the past, test kits were predominantly based on polyclonal antibodies. These are mixtures of antibodies recognizing different epitopes, each one, on the GH molecule. In this way, {\textit {in summa}}, all possible epitopes on the different GH-forms are more or less covered. Therefore, polyclonal antibodies may be expected to capture something like 'total-GH' in that sample.\par 

It is different, when monoclonal antibodies are used, as it is the case with most of the more contemporary assays. Uniformly, the same epitope is then recognized. Depending on the presence of this epitope exclusively on just one GH-form, or on many forms, the information content will be either of the kind 'total-GH' or 'particular GH form'. Enhanced specificity toward a particular form can be achieved if using two-step immunometric assays with two antibodies recognizing two different epitopes that are only present with that form.\par

All these antibodies, or antibody-mixtures, capture different fractions out of the spectrum of isoforms present in a given sample, according to the epitope(s) they recognize. Using different methods is equivalent, therefore, to allow for slightly different definitions of the measurand. This obviously accounts for a part of the variation of results if using different methods.\par

\begin{figure*}[]
\centering
\includegraphics[width=0.85\textwidth]{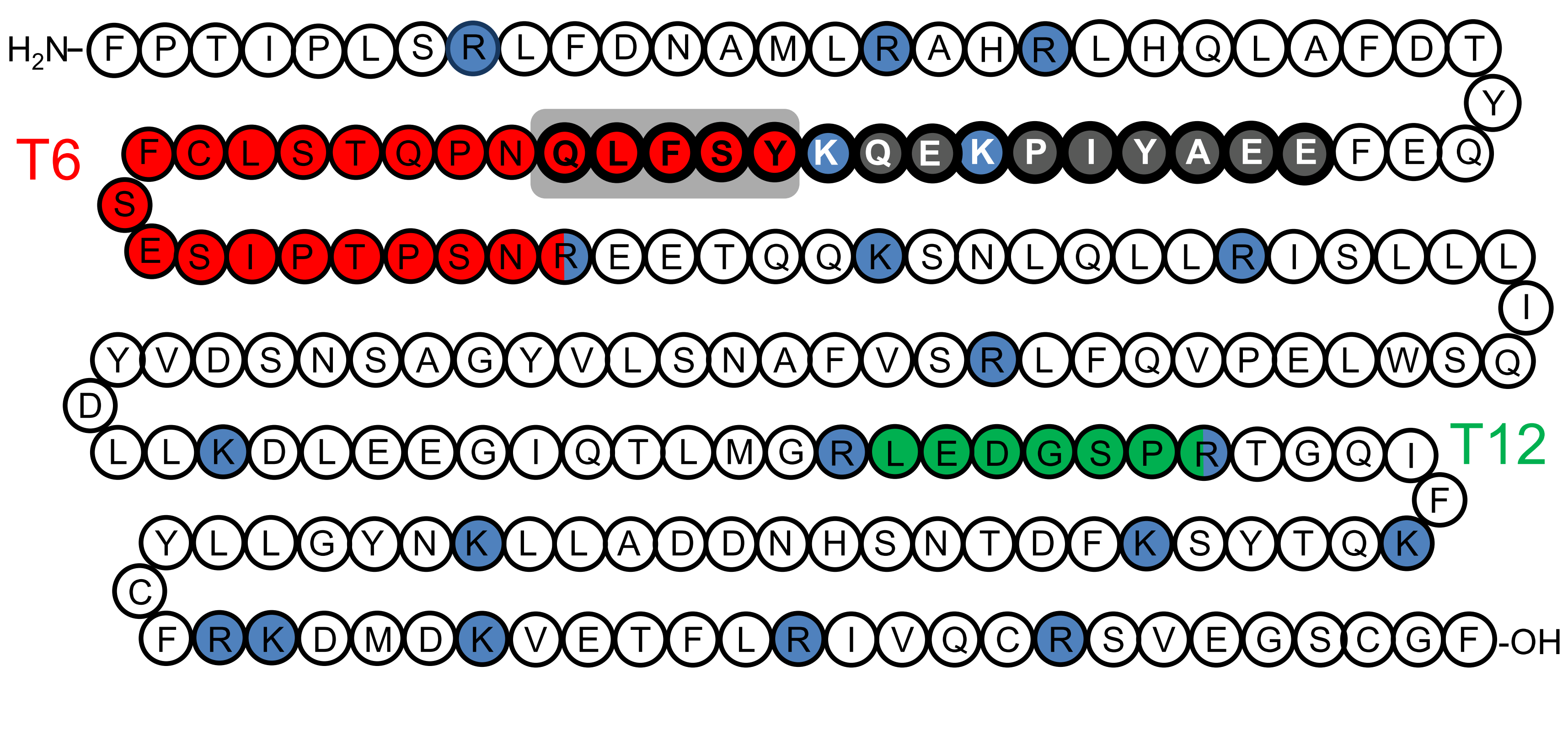}
\caption{Tryptic cleavage sites on 22~kDa-GH. Cleavage occurs after R and K, blue-filled circles. Substretches T6 and T12, used as signature peptides with the method illustrated in Figure~\ref{Workflow}, are highlighted in red and green, resp. ~T6 is present as shown, only with 22~kDa-GH,  while the 20~kDa isoform is missing the amino acids within the grey box (cp. Figure~\ref{GHSequence}). T12, on the other hand, will equally result from both forms. Therefore, T6 captures GH in an isoform-specific way, whereas T12 might be taken as measuring 'total-GH'.} \label{CleavageProducts} 
\end{figure*}

Potential cross-reactivities of antibodies constitute another noticeable input to inter-method divergences. Mistaking a non-GH antigen for GH, as it can happen owing to similarity of epitopes, may result in positive bias. However, it can equally well come to pass inversely. So, e.g., if in a sandwich-type assay just one of the two antibodies binds to that antigen, less of GH will be determined than what is present. One example for such antigen is pegvisomant, which may interfere with GH measurement if administered in therapy\cite{Paisley2007}.\par

However, pegvisomant is present only in sera of treated patients. Growth hormone binding protein (GHBP), which is generally present in serum samples at varying concentration, apparently constitutes still more of a problem. With commercial assays, GH-recovery has been reported to be reduced by up to about 40\% as a consequence of GHBP interference\cite{Boulo2013}. This may be taken as resulting from competition between the binding protein and the antibody for the same GH-epitope.\par

Apart from these uncertainties, which may be ascribed to variably specific recognition of the analytical target, a more elementary factor contributing to incompatibility has been that, originally, and for a long time, varying GH-preparations have been referred to as primary standards for assay-calibration. These were pituitary-tissue extracts containing the whole spectrum of isoforms in poorly characterized amount-ratios. Additionally aggravating it was, that measurement was linked to International Units of biological activity (IU), and not compatible, therefore, with the SI. This, however has been resolved by a WHO-agreement in 2001\cite{Bristow2001}, about using recombinant 22~kDa-GH (WHO-preparation 98/574) for calibration, from then on, as chemically well-defined pure primary standard. This, at the same time, allows for expression of the results in terms of the (molar) amount-of-substance, establishing a link to the SI, thereby.\par

Finally, non-commutability of kit-calibrators is discussed as another cause of inaccuracy. 'Commutability' addresses the requirement for the calibrator to behave in the same way as if it was a natural (patient) sample, though most of the time the calibrators are processed serum to which a known amount of GH had been spiked. Non-commutability, is attributed to presence of the analyte in a non-natural form in the calibrator, matrix-effects, and lack of specificity of measurement procedures\cite{Miller2006}, and will obviously affect the applicability of calibration to real samples, and compromise the SI-traceability of results.\par

\section*{Recognizing by mass}

\subsection*{Targeting signature peptides} 

Several MS approaches have been developed thus far for measurement of serum GH, on purposes to detect illicit administration in stock-breeding and horse-racing \cite{LeBreton2008,Bailly2008}, next to diagnosis of GH-related diseases in human medicine\cite{Arsene2010,Arsene2012,Pritchard2014}.\par

All these methods share in common, that enzymatic cleavage products of GH are quantified as surrogates (signature peptides)\cite{Geng2000}, in place of the protein itself. The preference of enzymatic fragments as analytical targets over the intact protein is characteristic for almost exclusively all methods presently being developed in quantitative proteomics. This  is obviously owing to the better compatibility with reversed phase chromatography and lower limits of detection/determination attainable with peptides if compared to the larger-sized precursor proteins. It is worth noting that, notwithstanding, the alternative of measuring the intact protein, as an option is kept in consideration too\cite{Ruan2011,Gucinski2012}. This appears mainly motivated by the complementary information, available in this way, about kind and abundance of different molecular variants of the protein.\par 

There are 20 tryptic cleavage points on 22~kDa-GH (i.e., after each K or R) as shown in Figure~\ref{CleavageProducts}. Of the resulting 21 substretches (products of exhaustive proteolytic cleavage), most can act as an analyte, provided that, they are specific enough in sequence, so as to be excluded from possibly being obtained similarly from any other protein present in the sample. \citeauthor{Arsene2010}\cite{Arsene2010,Arsene2012,Wagner2014} for instance, chose T6 and T12 to be quantified. Both sequences had been checked by comparison with the Swiss-Prot data base not to coincide with sequence stretches that might result from any other known human protein.\par

\subsection*{Integrating isotope-labeled standards} 

MS-Methods developed for clinical GH-measure\-ment almost compellingly have to employ isotope-labeled versions of the analyte as internal standards, so as to achieve the level of quantitative accuracy required. Ideally, the target protein is used to this, since closest mimicing the behavior of the analyte in both, extraction and proteolysis\cite{Brun2007,Lebert2011}. In the case of GH-measurement, recombinantly engineered 22~kDa-GH has been employed, either incorporating labeled nitrogen (U-$^{15}$N~GH)\cite{Arsene2012}, or labeled amino acids (K and R)\cite{Pritchard2014}.\par 

As alternative to whole-protein labeling, isotope labeled peptides (corresponding to the specified enzymatic fragments) have been shown to be applicable as internal standards too\cite{Arsene2010}. Using labeled peptides as internal standards, of course, takes as a prerequisite that, proteolysis will proceed to completion under the experimental conditions used (see \citeauthor{Arsene2008}\cite{Arsene2008} for a discussion). 

\subsection*{Extracting targeted peptides}

Once cleavage products have been specified, two fundamentally different strategies for generating and extracting them are conceivable: Promising it is, to first isolate the protein from the matrix and only then to enzymatically digest the analyte-enriched extract.\par

This rationale, referred to as 'protein level clean-up', is practiced with the method developments by \citeauthor{Bailly2008}\cite{Bailly2008} and \citeauthor{LeBreton2008}\cite{LeBreton2008}, who isolate GH by precipitating it from the serum at its iso-electric point, followed by solid-phase extraction, as well as by \citeauthor{Pritchard2014},\cite{Pritchard2014} who apply two-fold solid-phase extraction at different pH-values. Immunoaffinity-enrichment is likewise considered as an option for clean-up at the protein level\cite{Ackermann2007, Weiss2014}. This approach is also referred to as 'mass spectrometry based immunoassay' (MSIA), so as characterizing it as a hybrid of two philosophies pursued. However, with respect to the present purpose, employing an antibody-based extraction principle for GH ostensively would thwart the intent of developing a method that is completely independent of biases associated with analyte selection by epitope recognition.\par

\begin{figure}[H]
\includegraphics[width=0.45\textwidth]{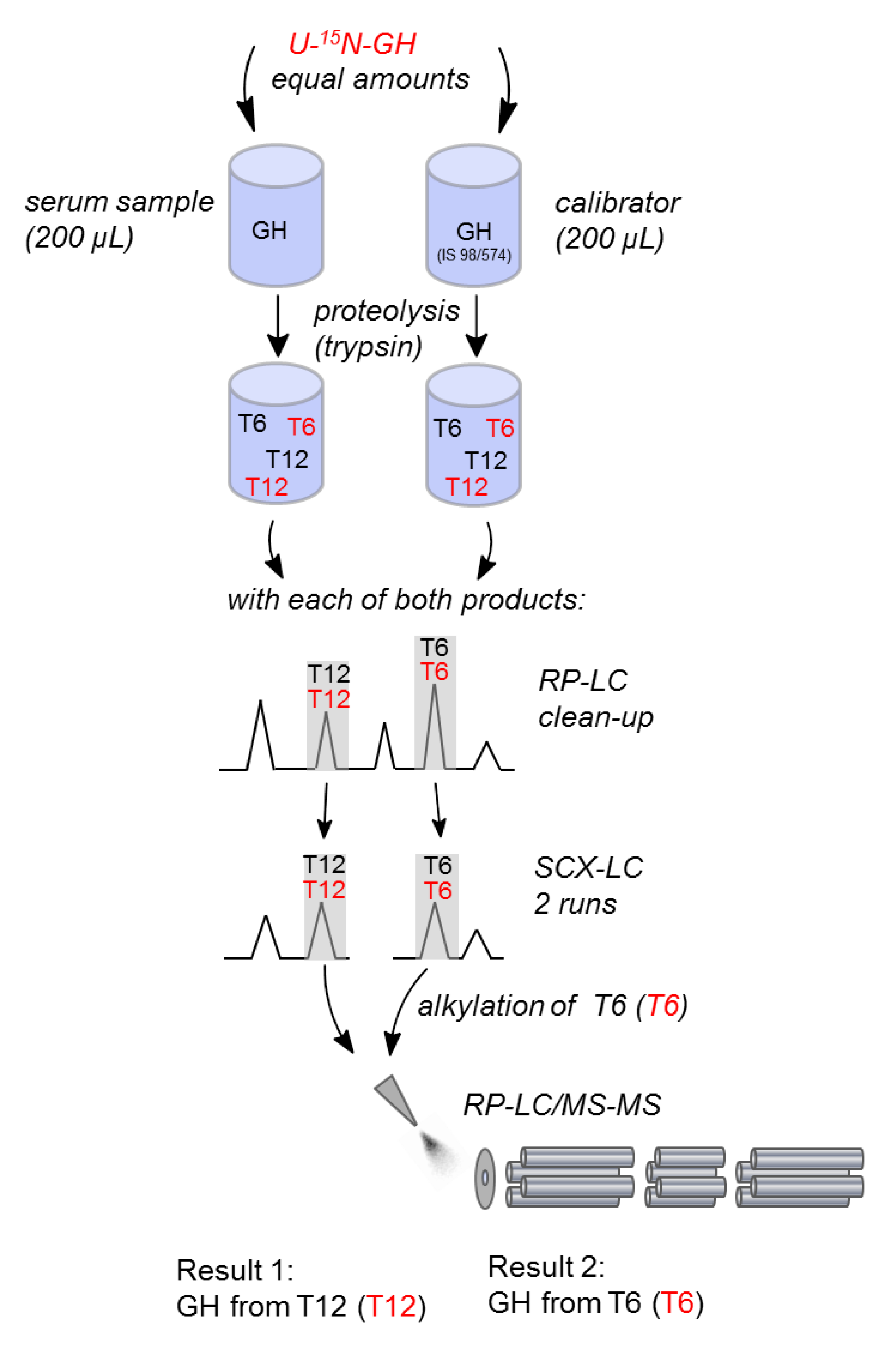}
\caption{Determination of GH by ID-MS using peptide level clean-up. The serum sample containing GH is enzymatically digested using trypsin as protease. Enzymatic GH-fragments, peptides T6 and T12, are quantified as surrogates in place of the intact protein. Along with each individual serum sample, a calibrator (serum plus a known amount of 22~kDa-GH, here: WHO~IS~98/574) is run and analyzed using the same procedure as with the sample. Isotopically labeled GH (here: U-$^{15}$N 22~kDa-GH) is spiked as the internal standard to 
both serum sample and calibrator at the beginning of the procedure. T6 and T12 given in red letters are to indicate the isotope-labeled versions of the fragments. (Reproduced with permission from \citeauthor{Wagner2014},\cite{Wagner2014}  
 {\em Eur J Endocrinol}, {\bf 2014}, accepted. \copyright 2014 European Society of Endocrinology)} \label{Workflow} 
\end{figure}

As opposed to these approaches, the alternative strategy (peptide level clean-up) appears to be considerably more challenging at first glance, indeed. In this case, the signature peptides have to be recovered after digestion from the serum sample as a whole, which results in a matrix of very high complexity. However, in practice, this has been proven to be manageable both, reliably stable quite as much, as reproducible. An outline of the implementation described by \citeauthor{Arsene2010}\cite{Wagner2014} is given in Figure~\ref{Workflow}. Data exemplifying accuracy and precision attainable with the type of method are reproduced in Table~\ref{MSPerformance}. Remarkably,  this appears to be the only MS-based method as yet, that has been proven to smoothly be applicable to patient sera within a clinical study\cite{Wagner2014}.\par

Reversed phase chromatography (RP-LC) followed by strong cation exchange (SCX) chromatography are used, in this case, for extraction of GH peptides T6 and T12 from the matrix. Smilar developments, targeting other proteins, successfully have been using antibodies instead, specifically developed for recognition of the signature peptides\cite{Anderson2004,Becker2012,Weiss2014}. This is more than likely to work quite as well with GH-quantification and may even be less time-consuming than chromatographic clean-up.\par

Using chromatography, on the other hand, obviates the need to develop a separate antibody against every targeted peptide, making it more easy to switch to another peptide, if required with a particular study.\par

\subsection*{Establishing SI-traceability}

Traceability can be obtained by reference to a calibrator consisting of (GH-stripped) blank serum, which is then fortified with a defined amount of recombinant 22 kDa-GH. The calibrator is run according to the same steps and conditions as applied with the sample. The analysis result, eventually, is obtained by relating the signal ratio (peptide/isotope-labeled peptide) measured for the sample, to the ratio obtained for the calibrator. In Figure~\ref{Workflow}, WHO IS 98/574 is specified as 22 kDa-GH material appropriate for the purpose. However, any other source would do quite as well, as long as it is of the correct amino acid sequence and sufficient in purity.\par 

Value-assignment to the calibrator stock solution is based on amino acid analysis (AAA), i.e., complete hydrolysis, quantitatively releasing the individual amino acids. These, in turn, are quantified by isotope dilution mass spectrometry (ID-MS). In this way, the amount of GH measured for a particular sample is traceable to the (molar) amount-of-substance of the constituent amino acids, which are simple molecules and readily available as purified reference standards. See \citeauthor{Burkitt2008}\cite{Burkitt2008} for a more detailed discussion of the concept.

\section*{Growth hormone by MS-- sco\-pe and bounds}

\subsection*{Analytical selectivity and sensitivity}

The promise of mass spectrometry consists in achieving enhanced selectivity by recognition of analyte mass, rather than -epitope, as principle for distinction from the background. Essentially, the data shown in Table~\ref{MSPerformance} account for accurate recovery of GH and indicate the absence of interferences. Apart from a bilateral comparison,\cite{Pritchard2014} no comprehensive interlaboratory study has been run yet for MS-measurement of GH. However, results obtained by \citeauthor{Cox2014}\cite{Cox2014} in a recent study about insulin-like growth factor~1 (IGF-1), which poses a similar analytical challenge, are promising.  IGF-1 is particularly interesting by its close physiological relationship with GH. In that study, five laboratories, using different LC/MS-MS platforms, demonstrated comparability within 5.6\%. U-$^{15}$N IGF-1 was employed as isotope labeled material in the example.

\begin{figure}[H]
\centering
\includegraphics[width=0.45\textwidth]{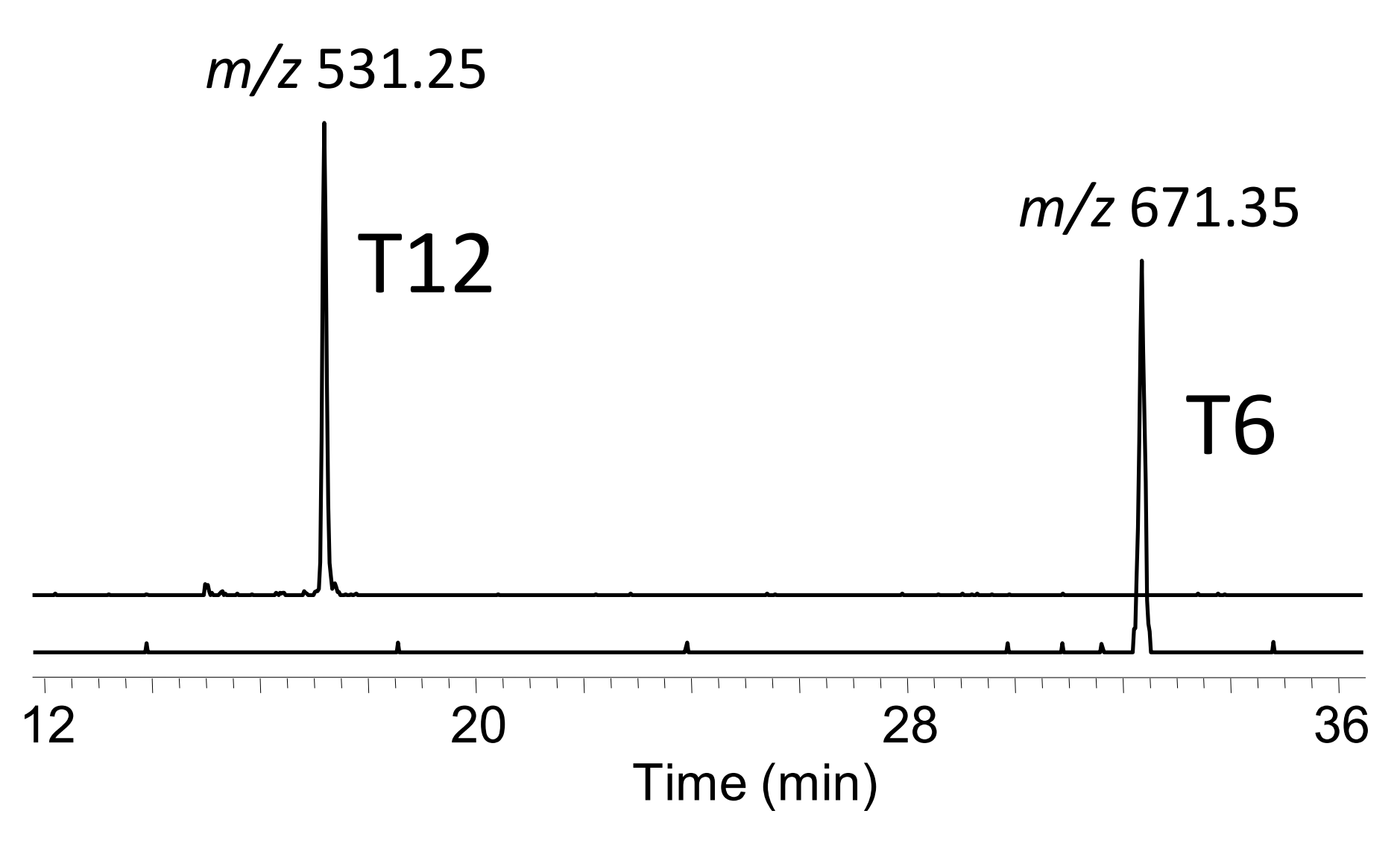}
\caption{Signals obtained from a serum sample at 1~ng/mL GH. Tryptic peptides T6 and T12 of GH had been extracted and were then analyzed by LC/MS-MS using the masses of collision fragments (at m/z 531.25 and 671.35) for monitoring. Instrument: Orbitrap-Elite at 60.000 mass-resolution. } \label{GHOrbitrap} 
\end{figure}

Selectivity and sensitivity achievable with MS is further illustrated by the example shown in Figure~\ref{GHOrbitrap}, with no other signals present in the ion chromatograms, but those at the known retention times of the peptides T6 and T12. In contrast to presently prevailing perception\cite{Kay2010,Vandenbroek2013} this example of a serum-GH measurement at 1~ng/mL also suggests that, MS-techniques can be run as sensitive, as making them competitive to contemporary antibody-based assays, even in concentration ranges required with glucose-tolerance tests, where cut-off levels have been reported of about 0.5~ng/mL.\cite{Bancos2013}\par 

\begin{table}[H]
\center
\begin{tabular}{cll}
& \multicolumn{2}{l}{Recovery (\%)}\\
\cline{2-3}
Spiked (ng/mL)  & By T6 & By T12\rule{0pt}{12pt}\\ 
\hline
4.63 & 103.5 & 104.1 \rule{0pt}{12pt}\\
4.58 & 97.2 & 100.4 \\
10.19 & 100.5 & 102.9 \\
10.25 & 98.9 & 99.0 \\
19.51 & 103.9 & 102.5 \\
19.32 & 102.0 & 96.8 \\
29.39 & 104.0 & 99.3 \\
30.31 & 103.1 & 100.4 \\
\hline
Mean recovery (\%) & 101.6 & 100.7 \rule{0pt}{12pt} \\
CV (\%) & 2.5 & 2.4 \\
LoQ (ng/mL) & 1.7 & 2.7
\end{tabular}
\caption{Recovery of 22~kDa-GH from serum by MS. (Reproduced in adapted form with permission from \citeauthor{Arsene2010},\cite{Arsene2010} {\em Anal Biochem}, {\bf 2010}, {\em 401}, 228--235 \copyright 2010 Elsevier Inc.)} \label{MSPerformance}
\end{table}

\begin{figure*}[ht]
\centering
\includegraphics[width=0.85\textwidth]{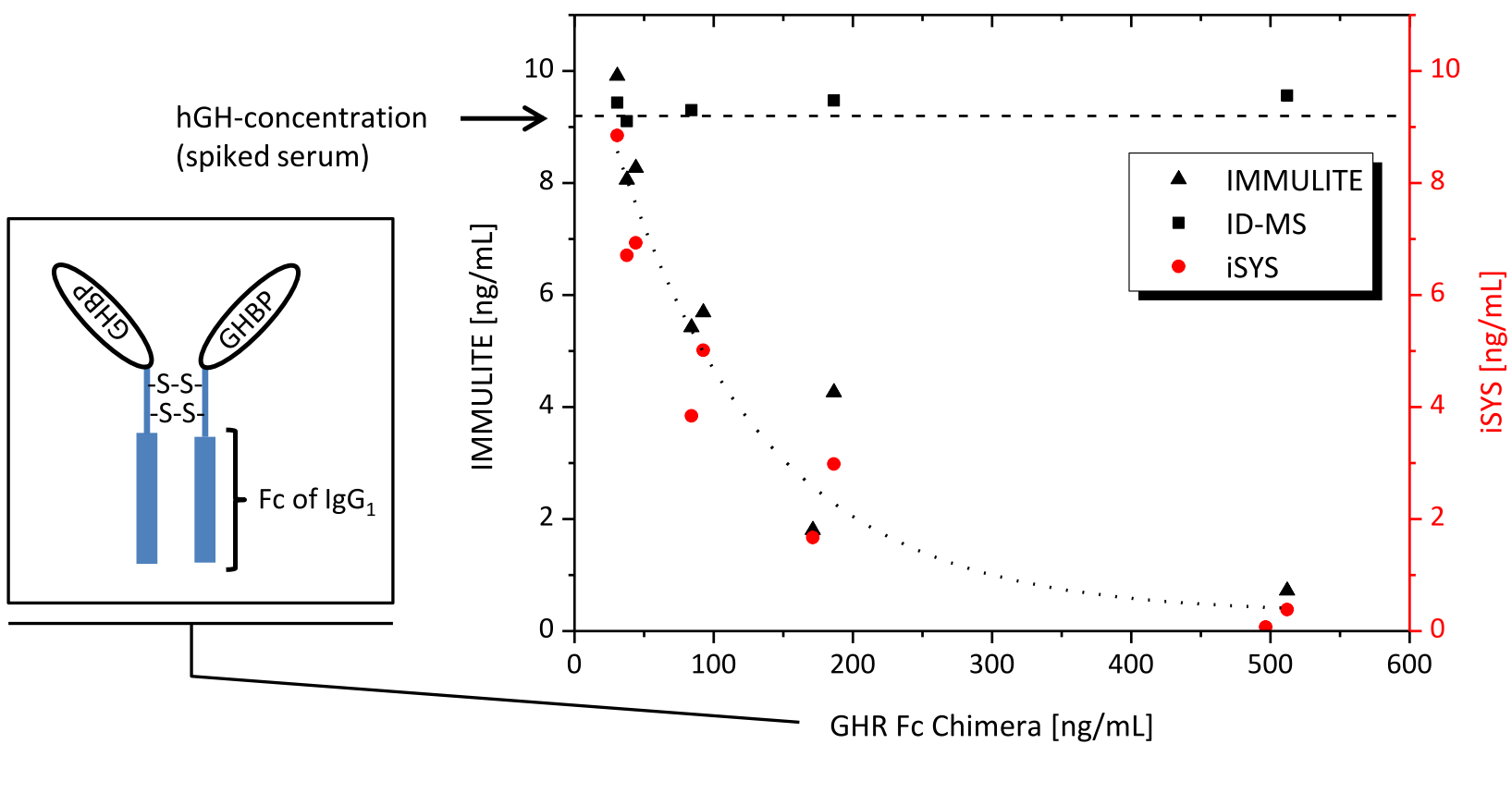}
\caption{Dependence of serum growth hormone (GH) assay results on binding protein (GHBP) level. Two commonly used antibody-based assays (IMMULITE, black triangles and iSYS, red circles) are compared with mass spectrometry (ID-MS, black squares). A GHR Fc chimera was spiked to serum so as to mimic increasing concentrations of the binding protein. The dashed horizontal line indicates the known concentration of 22~kDa-GH. ID-MS data shown are avarages from T6-based and T12-based GH measurement. (Reproduced from \citeauthor{Arsene2014a},\cite{Arsene2014a} {\em ArXiv e-prints}, {\bf 2014}, arXiv:1405.3895)} \label{GHBP} 
\end{figure*}

Needless to elaborate that, selectivity and sensitivity are largely interrelated and many factors in sample preparation, as well as instrumental set-up, have to be optimized so as to touch the limits of what is technically possible. The first version\cite{Arsene2010} of the protocol outlined in Figure~\ref{Workflow} (p.\pageref{Workflow}) was developed on a triple-quadrupole MS instrument. Monitoring collision fragments of T6 and T12, rather than unfragmented ions, essentially contributes to selectivity, as including a sequence-dependent parameter (fragment mass) for further distinction from matrix components, that might chance to have the same mass as T6 (or T12). The result shown in Figure~\ref{GHOrbitrap}, however, was obtained under conditions of high mass-resolution as achieved by using an Orbitrap mass spectrometer in place of the quadrupole instrument. This, apparently, does not just further increase selectivity, but it also improves the signal-to-noise ratio, and sensitivity, thus. Another option to enhance sensitivity consists in peptide conjugation to enhance amphipathicity, and ionization yield, by this way. So, e.g., based on selective tagging at cysteine, the limit of quantification (LoQ) could be improved by a factor of four in a later published variant of the protocol.\cite{Arsene2012}

\subsection*{Susceptibility to interfering proteins}

Owing to the advantage of high selectivity, as pointed out, mass spectrometric quantification of GH is not likely to be biased in presence of whatever background of accompanying proteins. Consequently, interference by growth hormone binding protein (GHBP), as commonly observed with many ligand-binding assays, is expected to be irrelevant with MS-based measurement.\par
This is confirmed by the data shown in Figure~\ref{GHBP}. Samples from a serum pool with defined concentration of GH (9~ng/mL) had been fortified with varying amounts of binding protein, such as reflecting the clinically relevant concentration range. The MS-method accurately recovers the total amount of GH, while the antibody-based assays report only the fraction that is accessible to the antibodies at the equilibrium of competitive binding.\par
Obviously, MS is highly reliable if it is about the total amount of GH in the sample. At the same time, it does not, of course, resolve in what aggregation state (free, receptor-bound, dimerized, etc.) the hormone is present. It cannot be ruled out, however, that these fractions contribute essential information to description of the physiological status of a patient and might evolve as relevant measurands too. \par

Taking account of this, \citeauthor{Frystyk2008}\cite{Frystyk2008} used ultrafiltration to remove the receptor-bound GH fraction prior to quantification of the remaining 'free GH'. A correlation was found, by these authors, between total and free GH for a given (constant) level of GHBP. This suggests redundance between the two measurands. If so, just one of the two would be neded for diagnostic purposes in reality, provided the GHBP level be known at the same time for that sample. Unfortunately, no MS-method, has been reported yet for GHBP-quantification, but could be useful in the context.\par

\citeauthor{Frystyk2008} also found an inverse correlation  of free-GH and GHBP levels (for a given total amount of GH). This is supportive of the notion that, GHBP serves to control the levels of 'active' GH available to the receptors. The example illustrates, that there may be no single and simple measurand sufficiently describing the clinically relevant interaction of the different molecular players in growth hormone signaling and metabolism.\par 

Returning on the high-selectivity advantage by MS, it should be mentioned that, GH-measurement by MS might cope with presence of pegvisomant, even in high excess, in a similar way, as it does with GHBP. This, of course, requires signature peptides to be selected, which are different between GH and pegvisomant.

\subsection*{Providing reference values}

MS-measurement, as pointed out, enables acquisition of SI-traceable values for (total) GH, which can serve either as reference values for this measurand within EQAS, or for calibration of LBAs.\par 

Viability of calibration of LBAs, concedingly, is limited by the fact that, the (unknown) GHBP level is involved in the outcome. Obviously, correct results can be only expected, if this level is the same in the calibrator as with the particular sample. This, however, is not the case, most of the time. As a strategy to eliminate this problem, dissociation of GHBP-antigen complexes under acidic conditions has been proposed to be applied to the samples prior to the assay\cite{Myler2010}.\par

Recently, Physikalisch-Technische Bundesanstalt (PTB) has started to provide MS-values for intercomparisons as shown in Figure~\ref{RfB} (p.\pageref{RfB}). PTB's capability of SI-traceable MS-based GH-measurement has been approved at international level by a 'Calibration and Measurement Capability (CMC-Claim)' registered at the Bureau International des Poids et Mesures (BIPM) [http://www.bipm.org/en/cipm-mra/].

\subsection*{Capturing isoform-specific}

Up to this point, discussion of the MS method was restricted by the assumption that, the 22 kDa-GH form exclusively appears in a sample. However, the concept of quantifying GH by signature peptides allows for separate measurement of the minor abundant GH-forms too, which are present, in a natural sample, next to 22~kDa-GH.\par

T6, for instance, as it is shown in Figure~\ref{CleavageProducts} (p.\pageref{CleavageProducts}), selectively captures the 22~kDa-GH form. The peptide cannot result from 20~kDa-GH, as the amino acids within the grey box would miss with this isoform (next to further deviations). For the purpose of quantification of the 20~kDa form, on the other hand, (20~kDa-) T4 would be one out of the possible signature peptides. (T4 refers to the fourth tryptic peptide produced from the protein, starting enumeration from the N-terminus.)\par

In a similar way, numerous peptides can be specified, which selectively measure a particular form of GH that is of interest in a study. In most of the cases, more than just one cleavage product is available to represent the same GH-form. If simultaneously quantifying more than one, the redundancy can be used to enhance reliability by averaging the results. Also important to observe that, the individual results will additionally provide 'checks and balances' as they should all be the same within measurement uncertainty. In the case of a discrepancy, particularly, if one predicted fragment should not be detected at all, this may indicate an unexpected structural deviation in the GH-sequence with that particular sample, altering the molecular mass of that peptide. Changes like this might result from e.g. mutation (substitution of amino acid) or side-chain modification. \par

\subsection*{Revising measurands}

The option of separately monitoring different GH-forms in high-selectivity/high-reliability mode will foreseeably impact progress in GH research too. Based on MS-measurement, decisions could become possible, where data from LBAs are leaving things ambiguous, presently.\par

One example, where such progress might be anticipated, is about the influence of the spectrum of circulating GH-isoforms on the occurrence of short stature in children. Different studies lead to different conclusions: While results by \citeauthor{Boguszewski1997}\cite{Boguszewski1997} suggested that elevated levels of non-22 kDa forms (which mainly is 20~kDa-GH) are  responsible for the (patho-) physiological condition, \citeauthor{Tsushima1999}\cite{Tsushima1999} reported constant ratios of 20 kDa- and 22 kDa GH, regardless of normal or short stature. Also, it was proposed\cite{Wallace2001} that, minor isoforms (neither 22~kDa, nor 20~kDa) might account for the effects observed by \citeauthor{Boguszewski1997}\par
An MS-based study may be expected to reduce ambiguity by providing better analytical resolution, as enabled by mass-selective differentiation of isoform-specific peptides.

\begin{figure}[]
\hfill
\includegraphics[width=0.45\textwidth]{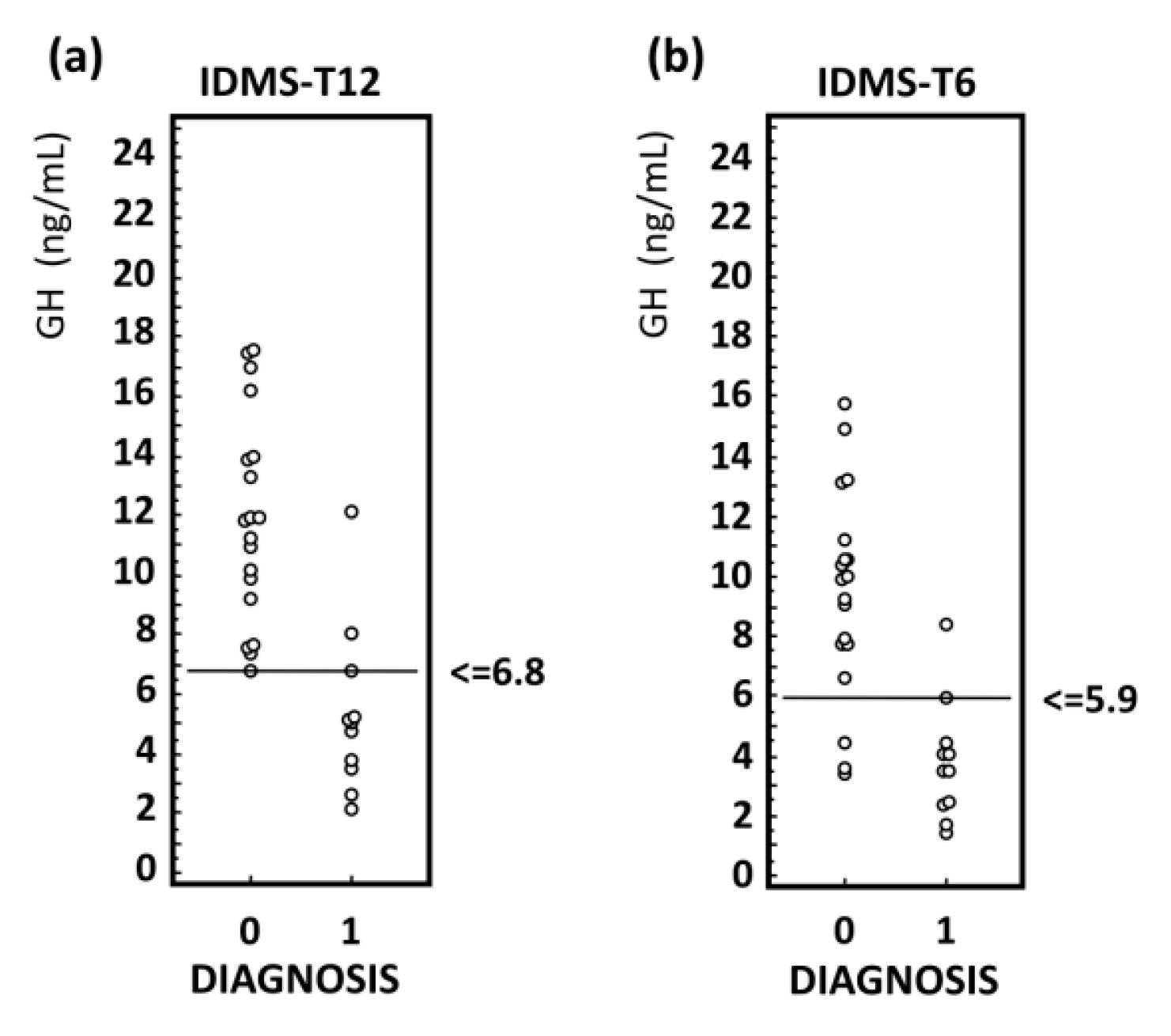}
\caption{MS in definition of diagnostic cut-off levels. Results are shown for healthy (0) and GH-deficient children (1). Signature peptides 22~kDa-T12 (a) and -T6 (b) were quantified. Monitoring T12 is equivalent to mapping ‘total GH’ whereas T6 reflects 22 kDa-GH. (Graphical representation of data contributed to a recent publication: \citeauthor{Wagner2014}\cite{Wagner2014} {\em Eur J Endocrinol}, {\bf 2014}, accepted)}\label{CutOff} 
\end{figure}

\subsection*{Defining cut-off levels}

The determination of diagnostic decision levels generally involves lengthy and elaborate clinical studies. It is particularly important therefore, to have the measurand well-defined at the outset. As it has been pointed out, present knowledge about what is the optimal measurand in GH-related diseases (sum of GH-forms, individual isoform(s), bound/free ratio, other?) may be advanced by MS-methods of measurement. \par

A major limitation associated with exclusive reference to LBA measurement in determination of cut-off values it is that, the antibody may not be available anymore at a later point, as well as the calibrator may be replaced by another one, which changes will hamper applicability (reproducibility) then, of~results.\par

MS, as opposed to this, has the potential of establishing sustainable reproducibility, independent of what laboratory or MS-platform being used. An example for MS-based cut-off determination for diagnosis of GH-deficiency is given in Figure~\ref{CutOff}. In spite of the limited size of the data set ($n=30$ patients), it appears that both, T6 and T12, equally discriminate between healthy and GH-deficient children. The ratio between the cut-offs obtained for T6 and T12, 5.9 and 6.8 ng/mL, is roughly reflecting the expected fraction of 22 kDa GH within total GH (here: 87\%). In the original study,\cite{Wagner2014} the purpose of including MS-measurement was establishing a metrological link to the LBA-based cut-off values.

\subsection*{Supporting doping control}

Present detection strategies for GH-doping in sports have been reviewed by \citeauthor{Baumann2012}\cite{Baumann2012}. The particular potential of MS-based proteomics methods for detection of protein doping has been highlighted by \citeauthor{Kay2010},\cite{Kay2010} and a review, summarizing recent developments in the area, has been published by \citeauthor{Vandenbroek2013} \cite{Vandenbroek2013} Insufficient sensitivity uniformly was discussed to be the main obstacle to application of MS. However, the examples and data provided in the previous sections indicate that, the sensitivity needed may be attainable if applying the latest generation of MS-instruments in combination with optimized sample preparation.\par
In growth hormone doping, recombinant 22~kDa-GH, putatively, is the common form of administration. The intake will increase the serum GH-level for about 20~h (window of opportunity in doping-control). Nevertheless, definition of an (absolute) cut-off concentration, as a way to unambiguously detect such event, is not viable: Fluctuations from as little as 0.01 ng/mL up to about 20-30 ng/mL (or even more) are not uncommon, as resulting from just normal biorhythm, with a peak-to-peak interval of 2--3~h. In addition, elevated GH-levels may also occur as response to physical exercise, stress or fasting.\cite{Baumann2012}.\par

\begin{figure}[]
\hfill
\includegraphics[width=0.45\textwidth]{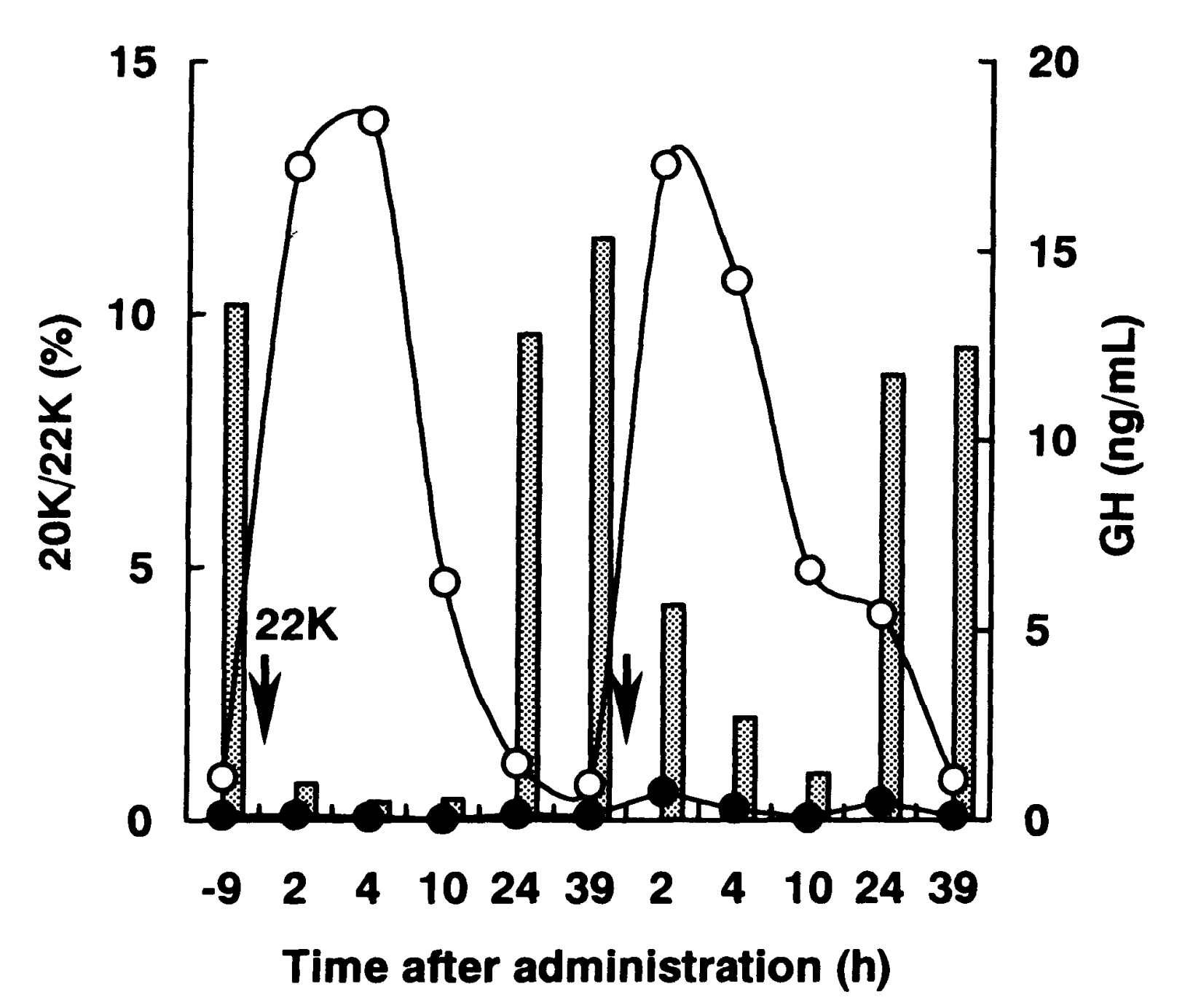}
\caption{Changes in serum concentrations of 20~kDa-GH ($\bullet$) and
22~kDa-GH ($\circ$) and the ratio of 20~kDa/22~kDa (bar) after administration
of 22~kDa-GH. The arrows represent the time of administration.
(Reproduced with permission from \citeauthor{Momomura2000}\cite{Momomura2000}, {\em Endocr J}, {\bf 2000}, {\em 47}, 97--101~\copyright 2000 The Japan Endocrine Society)}\label{DopingTestExample} 
\end{figure}

Therefore, 'isoform-differential' tests have been developed which do not depend on absolute concentrations of the hormone. Instead, the disturbance of the natural spectrum of isoforms, which results from exogenous administration of the 22~kDa-GH, is used as an indicator. Based on this notion, e.g. \citeauthor{Wu1999}\cite{Wu1999} and \citeauthor{Bidlingmaier2009}\cite{Bidlingmaier2009} developed methods using a combination of antibodies recognizing different GH-fractions, each: One type of antibodies specifically measures only 22~kDa-GH ('rec GH'), while the other one recognizes all forms secreted by the pituitary ('pit GH'). Elevated rec/pit-ratios are taken indicative of doping with 22~kDa-GH.~ \citeauthor{Momomura2000},\cite{Momomura2000} on the other hand, proposed a combination of antibodies, which both are isoform-specific, targeting the 22~kDa- and the 20~kDa-form, respectively. The principle is illustrated in Figure~\ref{DopingTestExample}.\par

Exploration of the potential of MS in this area is suggesting itself. MS-based acquisition of isoform-ratios could be expected to either directly improve discrimination between both classes of samples (doped or not), or, at least, be used to back-up findings by LBAs. The epitope-recognition principles of the existing tests can be translated into signature peptides for MS-measurement in a straightforward way: T6/T12-ratios, e.g. may be deemed to map a quantity corresponding to the rec/pit assay, as T6 selectively measures 22~kDa-GH, while T12 represents 'total GH' (see Figure~\ref{CleavageProducts}). Replacing T12 by T4 (confined to the substretch exclusively present on 20~kDa-GH), measuring T6/T4-ratios thus, would constitute an MS-analogy to the method of \citeauthor{Momomura2000}\cite{Momomura2000}\par

\section*{Future perspective}

The previous sections support the conclusion that, MS provides an effective alternative to traditional (antibody-based) ligand-binding assays in GH-measurement. Advantages of MS result from {\em(i)} high analytical selectivity enabled by mass-recognition instead of epitopes, {\em(ii)} flexibility and precision in targeting particular (iso-) forms by using signature peptides and {\em(iii)} ease of integration of isotope labeled internal standards into the protocol. By virtue of these features, many of the problems encountered with ligand binding assays either do not occur at all with MS-methods, or may be overcome in a simple way. This does not just apply to interferences with binding- or other proteins, and commutability issues resulting therefrom, but also to imprecisions in definition/recognition of the targeted epitope or GH-form, in a more general sense.\par

The technology has matured to the extent as being available for either direct measurement of patient samples (thus replacing ligand-binding assays), or value-assignment to serum-based control- and reference materials (supporting calibration and proficiency testing of antibody-based, ligand-binding assays). Harmonizing efforts for present commercial tests, no doubt, can capitalize on comparison with MS-results. Equally important it is, however, that the option of highly selective (iso-) form specific GH-quantification by MS has much promise as analytical tool to support further investigation into the putatively different biological functions of these forms, and if different, which particular functions these are.\par

The advantages associated with MS- based measurement  eventually might outweigh the main assets of ligand-binding assays, as low costs and ease of use in routine. The more so, if considering that, results are scarcely needed within a few hours in diagnosis of GH-related disorders. The time, MS analysis takes, and higher costs per analysis, in practice might pay off in terms of more accurate classification of patients, and of costs avoided, which would result from inappropriate treatment. Therefore, the authors speculate, MS even might supersede the antibody-based assays in clinical practice, within a few years' time.

\footnotesize
\bibliography{MSGH}

\end{document}